\documentclass[english,aps,prb,preprint,superscriptaddress,showpacs,showkeys]{article}
\usepackage[T1]{fontenc}
\usepackage[latin9]{inputenc}
\usepackage{amsmath}
\usepackage{amssymb}
\usepackage{graphicx}

\makeatletter

\usepackage{cite}

\usepackage{amsfonts}

\usepackage{times}

\usepackage{epstopdf}

\@ifundefined{definecolor}{\usepackage{color}}{}

\usepackage{psboxit}
\newcommand{\wccComment}[1]{
\ifthenelse{\boolean{printcomments}}
{ \hfill\\
\begin{minipage}{\textwidth}
\begin{boxitpara}{box 0.95 setgray fill}
\textcolor{blue}
{{\sf WCC:} \sf \hspace{0.25in} #1}
\textcolor{black}
\end{boxitpara}
\end{minipage}\hfill\\
}
{}<++>
}

\newcommand{\wccAddition}[1]{\ifthenelse{\boolean{printcomments}} 
{\textcolor{blue}{#1}\textcolor{black}}
{}
}

 \usepackage{epstopdf}\usepackage{times}\textwidth16cm \textheight22cm \evensidemargin0.3cm
\usepackage{babel}

\usepackage{babel}

\usepackage{babel}

\makeatother

\usepackage{babel}
\begin{document}

\title{Capillary instability in nanowire geometries}

\author{T. Frolov$^{1}$, W. C. Carter$^{2}$ and M. Asta$^{1}$}

\maketitle
$^{1}$ Department of Materials Science and Engineering, University
of California, Berkeley, California 94720, USA

$^{2}$ Department of Materials Science and Engineering, Massachusetts
Institute of Technology, Cambridge, MA 02139, USA

\textbf{The vapor-liquid-solid (VLS) mechanism~\cite{wagner:89}
has been applied extensively as a framework for growing single-crystal
semiconductor nanowires for applications spanning optoelectronic,
sensor and energy-related technologies~\cite{springerlink:10.1007/s10853-012-6388-0,Cui17082001}.
Recent experiments have demonstrated that subtle changes in VLS growth
conditions produce a diversity of nanowire morphologies, and result
in intricate kinked structures that may yield novel properties~\cite{TersoffNANO11,ShadiNANO11,doi:10.1021/nl902013g,Lieber,0034-4885-73-11-114501}.
These observations have motivated modeling studies that have linked
kinking phenomena to processes at the triple line between vapor, liquid
and solid phases that cause spontaneous {}``tilting\textquotedblright{}
of the growth direction. Here we present atomistic simulations and
theoretical analyses that reveal a tilting instability that is intrinsic
to nanowire geometries, even in the absence of pronounced anisotropies
in solid-liquid interface properties. The analysis produces a very
simple conclusion: the transition between axisymmetric and tilted
triple lines is shown to occur when the triple line geometry satisfies
Young's force-balance condition. The intrinsic nature of the instability
may have broad implications for the design of experimental strategies
for controlled growth of crystalline nanowires with complex geometries.}

Recent experimental and modeling studies have expanded the theoretical
understanding of the thermodynamic and kinetic factors underlying
the process of VLS growth, and the origins of observed changes in
growth morphologies. In this context, the modeling studies have investigated
the geometry of wetting of liquid droplets on faceted nanowires, and
have demonstrated transitions between axisymmetric and non-axisymmetric
wetting configurations~\cite{schwalbach:024302,JMR:8368023,Carroll1984195,roper:114320,muralidharan:114305}.
It has been shown that liquid droplet de-pinning from the corners
of a rigid nanowire occurs simultaneously with the onset of a tilted
(or, kinked) growth morphology~\cite{muralidharan:114305}. Experimental
studies of kink formation in VLS growth have demonstrated that such
tilting is accompanied by changes in the orientation and shape of
the solid-liquid interface~\cite{doi:10.1021/nl902013g}, and such
observations have been reproduced in computational models of nanowire
growth that assume highly anisotropic solid-liquid interface properties~\cite{PhysRevLett.102.206101,PhysRevLett.107.265502,doi:10.1021/nl1027815}.
While these continuum models have succeeded in reproducing the key
features of kinking and tilting processes in VLS growth, they rely
on assumptions about processes at the vapor-liquid-solid triple line
that are difficult to verify directly from experimental observations.
Furthermore, it is difficult to distill a simple explanation that
directly conveys to process design and control.

In this paper, we report the results of three-dimensional molecular-dynamics
(MD) simulations for a model system consisting of a diamond cubic
solid with \{211\} side-walls and a {[}111{]} nanowire axis wetted
by a liquid droplet in a VLS geometry (see Methods section for details).
In addition, we produce a model that results in a simple physical
explanation for the onset of tilting.

\begin{figure}
\begin{centering}
\includegraphics[width=1\textwidth]{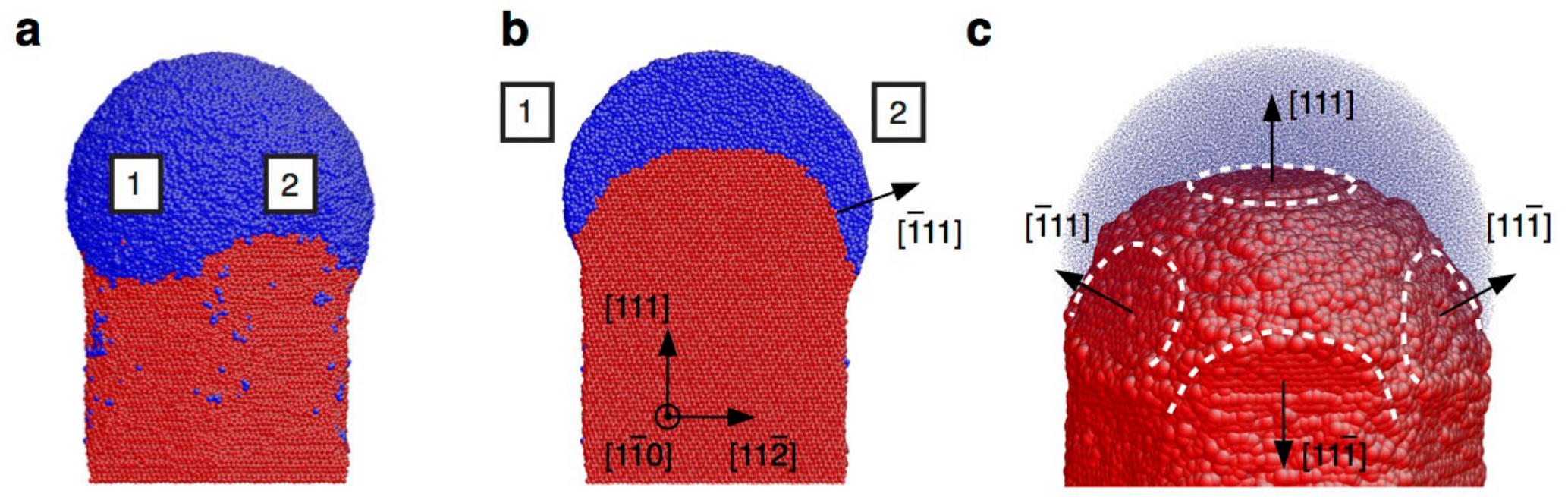}
\par\end{centering}

\caption{\textbf{Equilibrium wetting configurations of simulated nanowires.}
\textbf{a,} Nanowire with solid (colored red) and liquid (blue) phases
as identified according to a local crystalline order parameter. The
solid-liquid-vapor contact line has a variable elevation as it traverses
the perimeter of the nanowire: this triple line dips below the facets
with $[\overline{1}11]$ and $[11\overline{1}]$ normals, and above
the facet with $[1\overline{1}1]$ normal. Slice \textbf{b} shows
the structure of the solid-liquid interface. \textbf{c,} A view of
the solid liquid interface with semitransparent liquid phase illustrating
its faceted nature. \label{fig:Equilibrium-wetting}}
\end{figure}

Figure~\ref{fig:Equilibrium-wetting} shows a snapshot from the current
MD simulations illustrating an equilibrated nanowire configuration.
The simulations reveal a complex geometry associated with the wetting
of the crystal surfaces. Figure~\ref{fig:Equilibrium-wetting}a demonstrates
that the contact line (i.e., the solid, liquid and vapor (vacuum)
phases' triple line) dips below one facet indicated by $\boxed{2}$
and climbs above the neighboring facet labeled as $\boxed{1}$. These
triple lines on facets $\boxed{1}$ and $\boxed{2}$ are formed by
interfaces with different crystallographic orientations (c.f., Figure
\ref{fig:Equilibrium-wetting}b). The anisotropy of interface free
energies is known to produce such triple line {}``rumpling''.\cite{Zucker}

The 3d geometry of the solid-liquid interface is illustrated in Figure~\ref{fig:Equilibrium-wetting}c
which features four $\{111\}$ faceted orientations separated by interfaces
that appear atomically rough in nature. The figure also shows the
presence of three $\{111\}$ solid-vapor surfaces adjacent to the
{}``raised'' contact lines. These facets are qualitatively similar
to the sawtooth geometries commonly reported for VLS growth of Si
nanowires using Au catalysts.\cite{PhysRevLett.95.146104}

The wetting geometry shown in Figure~\ref{fig:Equilibrium-wetting}
involves contact angles that vary with the volume of the liquid ($V_{l}$),
as illustrated in Figure \ref{fig:Kinking_MD}. It is evident from
the figure that the angle between solid-liquid and vapor-liquid interfaces
($\phi_{vs}$) increases with increasing size of the liquid droplet.
We compare the nanowire contact angles $\phi_{vs}$ and $\phi_{lv}$
with equilibrium wetting angles of an isolated liquid droplet on a
\{111\} surface, $\phi_{vs}^{Young}$ and $\phi_{lv}^{Young}$, shown
in Figure \ref{fig:Kinking_MD}c in blue and magenta, respectively.
We find that for the range of liquid volumes where $\phi_{vs}$ is
less than $\phi_{vs}^{Young}$, the symmetric configurations are observed
to remain stable in the simulations. However, as nanowire contact
angles approach the Young's values with increasing liquid volume,
we observe a spontaneous tilting of the droplet to the side of the
nanowire. We were unable to observe stable equilibrium configurations
with $\phi_{vs}>\phi_{vs}^{Young}$. The model presented below yields
the same result, but for a simplified geometry.

\begin{figure}
\begin{centering}
\includegraphics[height=0.5\textheight]{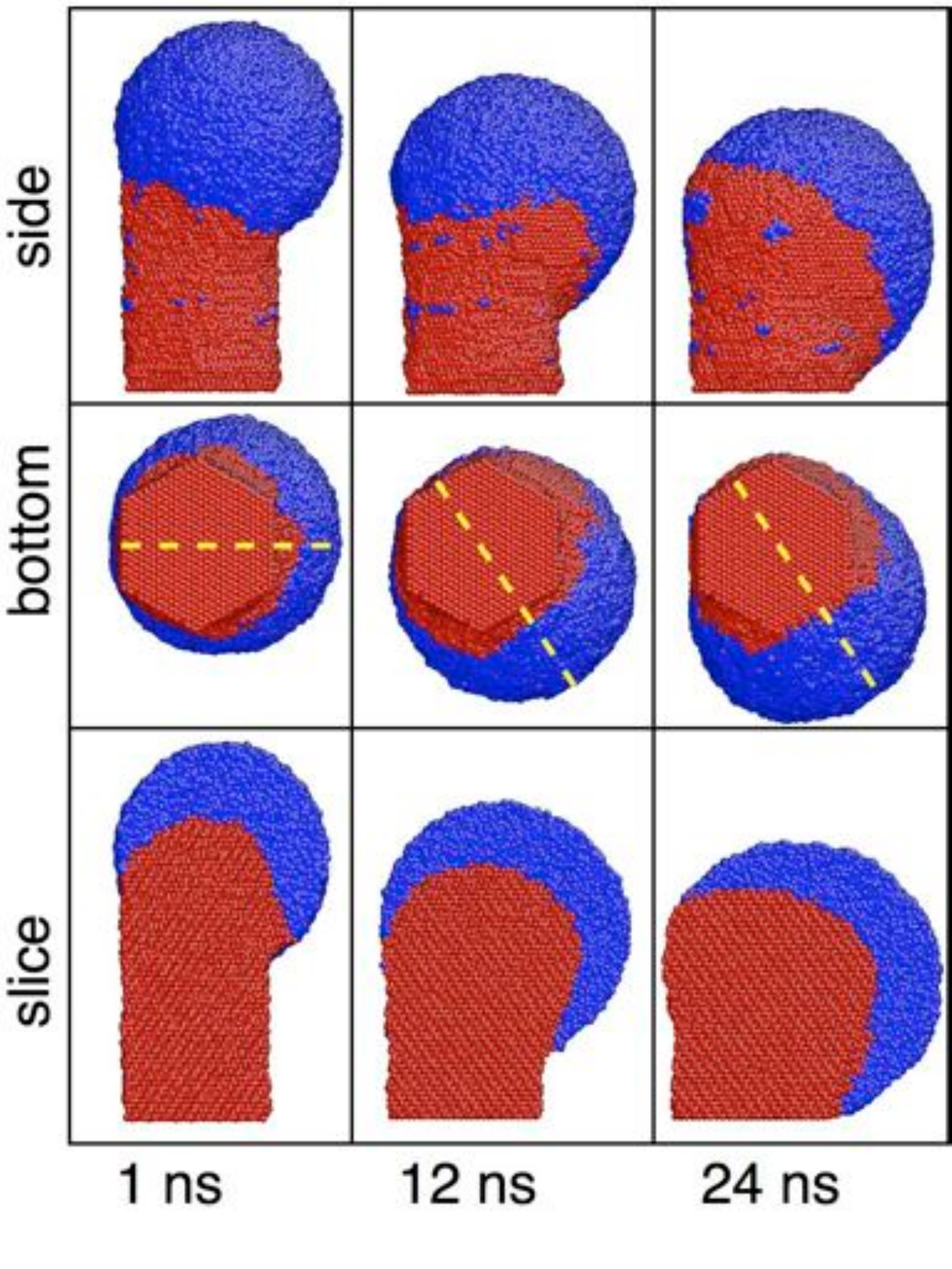}
\par\end{centering}

\caption{\textbf{Transition from symmetric to tilted configurations.} Time
sequence of snapshots demonstrating the evolution of the droplet tilting
during a 24 ns MD simulation. Top and middle panels show side and
bottom views of the nanowire. The middle panel illustrates that the
droplet tilting direction changes with time. The bottom panel shows
slices going through the nanowire along planes marked in the middle
panel by yellow dashed lines. Initially the tilting of the droplet
leads to the formation of a large $[\overline{1}11]$ solid-liquid
interface facet. At later times the droplet wets the side wall of
the nanowire, with the solid-liquid interface becoming more rounded.\label{fig:Instability-MD}}
\end{figure}

Figure~\ref{fig:Instability-MD} shows the nature of the tilting
process. The initial step involves tilting towards the $[11\overline{2}]$
direction, forming a large $(\overline{1}11)$ solid-liquid interface
facet. This observation is in qualitative agreement with experiments\cite{doi:10.1021/nl902013g}
and previous mesoscale simulations. However, as time progresses the
droplet does not lock into a new equilibrium wetting configuration,
but rather continues to advance along the nanowire surface (c.f.,
top panels of Figure~\ref{fig:Instability-MD}), while the center
of mass of the droplet rotates towards a different facet (c.f., middle
panels). During this process the shape of the solid-liquid interface
evolves on the same time scale as shown in the bottom panel of Figure~\ref{fig:Instability-MD}.
Specifically, during the process the solid part of the nanowire covered
by liquid increases and becomes more spherical. The top part of the
nanowire remains wetted in this process and the unpinning of the droplet
never occurs in the simulation. This behavior is qualitatively distinct
from the assumptions underlying recent models that consider wetting
transitions on fixed solid shapes~\cite{muralidharan:114305}.

To further explore the correlation between the tilting instability
and droplet volume we performed a separate simulation in which we
started with a tilted configuration and removed part of the liquid,
thus reducing the droplet volume. In the subsequent equilibration,
the tilted configuration gradually returned to the symmetrical geometry
with the liquid droplet mounted on top of the nanowire as in Figure
\ref{fig:Equilibrium-wetting}. This convincingly demonstrates that
the symmetric wetting configuration is an equilibrium thermodynamic
state for the smaller droplet volumes, and there is a critical volume
at which a transition to a non-symmetric configuration occurs.

\begin{figure}
\begin{centering}
\includegraphics[height=0.7\textheight]{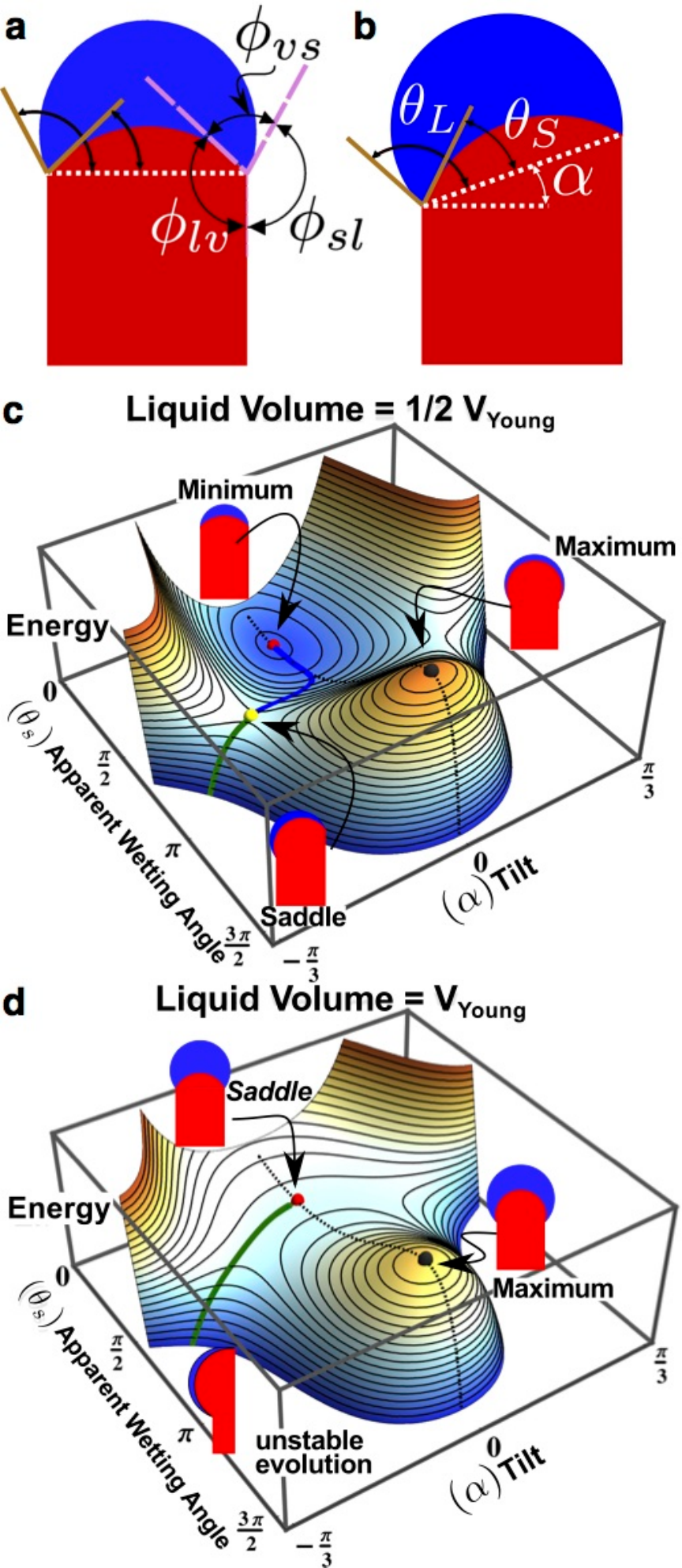}
\par\end{centering}

\caption{\textbf{Two Dimensional Model of Nanowire Capillary Instability.}
\textbf{a,} The angles, $\phi$ , which vary with liquid volume. \textbf{b,}
The model parameters that represent the degrees of freedom for capillary
shapes: the asymmetric tilt angle $\alpha$ and the apparent wetting
angle $\theta_{s}$ appear as coordinates in the free energy surfaces
for different liquid volumes in the middle and bottom figure. For
a symmetrical equilibrium configuration ($\alpha=0$) $\phi_{vs}=\theta_{L}$
and $\phi_{lv}=\theta_{s}+\pi$/2. The free-energy surfaces are plotted
for interface energy ratios $\gamma_{lv}/\gamma_{vs}=9/10$ and $\gamma_{sl}/\gamma_{vs}=3/10$.
\textbf{c,} Free-energy surface plotted for a volume which is half
of the Young's volume (i.e., the volume at which the $\phi_{ij}$
satisfy Young's equation). When $V_{l}<V_{l}^{Young}$ , there is
a metastable minimum at a finite $\theta_{s}$ and zero tilt angle
($\alpha=0$), and symmetric saddle points for $\alpha\neq0$ . The
blue and green lines were computed by steepest descent and represent
the most probable path for the system to transform from a symmetric
nanowire wetting geometry to a tilted configuration. \textbf{d,} As
the volume increases, the saddle and metastable points approach each
other and converge when $V_{l}=V_{l}^{Young}$. For liquid volumes
larger than $V_{l}^{Young}$ the metastable minimum corresponding
to a symmetric nanowire geometry disappears, and symmetric configurations
are unstable with respect to tilting.\label{fig:Two-Dimensional-Model}}
\end{figure}

The observations above motivate the development of a wetting model
to provide further insights into the nature of the tilting instability.
We consider the 2d nanowire geometry illustrated in Figure $\ref{fig:Two-Dimensional-Model}$a,b.
Liquid-vapor and solid-liquid interfaces are assumed to be isotropic
and therefore have circular shapes. The interfaces are connected at
VLS triple junctions located on the side wall of the nanowire. We
assume that the solid-liquid interface is mobile and can adjust its
shape to minimize energy, while the VLS triple junctions are restricted
to move along the side walls. The wetting geometry is fully specified
by a tilting angle $\alpha$ and two apparent wetting angles $\theta_{s}$
and $\theta_{l}$ defined in Figure~\ref{fig:Two-Dimensional-Model}b.
The total surface free energy $F^{s}$ of the system is given as 
\begin{equation}
F^{s}=\gamma_{lv}L_{lv}+\gamma_{sl}L_{sl}+2\gamma_{vs}L_{vs}\label{eq:F_gammaL}
\end{equation}
 where $\gamma_{lv}$, $\gamma_{sl}$ and $\gamma_{vs}$ are liquid-vapor,
solid-liquid and vapor-solid interface free energies, respectively
and $L_{lv}$, $L_{sl}$ and $2L_{vs}$ are the lengths of these interfaces.

We recognize that a two-dimensional fully isotropic treatment is a
simplified model. However, the results that we obtain below will be
sufficiently general to justify such an approximation. We will continue
to use a three-dimensional vocabulary volume and area to represent
their analogs in two dimensions: area and length.

To find the equilibrium wetting configurations we consider variations
of shape subject to the constraints that the total volumes of the
liquid and solid phases are constant 
\begin{equation}
V_{s}=L_{vs}+\frac{R_{sl}^{2}}{2}\left(2\theta_{s}-\sin\left(2\theta_{s}\right)\right)\label{eq:constraint1solid}
\end{equation}

\begin{equation}
V_{l}=\frac{R_{lv}^{2}}{2}\left(2\theta_{l}-\sin\left(2\theta_{l}\right)\right)-\frac{R_{sl}^{2}}{2}\left(2\theta_{s}-\sin\left(2\theta_{l}\right)\right)\label{eq:constraint2liquid}
\end{equation}
 where $R_{lv}$ and $R_{sl}$ are the radii of curvature of the liquid-vapor
and solid- liquid interfaces, respectively. By means of Eqs. ($\ref{eq:constraint1solid}$)
and ($\ref{eq:constraint2liquid}$), two variables from Eq. (\ref{eq:F_gammaL})
can be eliminated, resulting in $F=F(\theta_{s},\alpha,V_{l,}\left(\frac{\gamma_{lv}}{\gamma_{vs}}\right),\left(\frac{\gamma_{sl}}{\gamma_{vs}}\right))$.
At a given set of isotropic surface tensions (i.e. the ratios $\left(\frac{\gamma_{lv}}{\gamma_{vs}}\right)$
and $\left(\frac{\gamma_{sl}}{\gamma_{vs}}\right)$) and fixed volume
of the liquid $V_{l}$, the total surface free energy is a function
of the apparent contact angle $\theta_{s}$ and the tilting angle
$\alpha$. The equilibrium configurations correspond to states with
$\frac{\partial F}{\partial\alpha}=0$ and $\frac{\partial F}{\partial\theta_{s}}=0$.
From these two equations, it can be derived that the following conditions
are satisfied at equilibrium:

\begin{equation}
\gamma_{vs}=\gamma_{sl}\cos\left(\dfrac{\pi}{2}-\theta_{s}\pm\alpha\right)+\gamma_{lv}\cos\left(\theta_{l}\pm\alpha-\dfrac{\pi}{2}\right)\label{eq:force_balance}
\end{equation}
 Eq. (\ref{eq:force_balance}) can be interpreted as a balance of
capillary forces at triple junctions in the direction parallel to
the sidewalls of the nanowire, with plus and minus sign referring
to each of the two triple junctions. The condition (\ref{eq:force_balance})
is expected: because TJs parallel motion is not constrained, in equilibrium
the net capillary force along this direction must be zero. The forces
are not zero normal to the nanowire surface because the triple junction
is pinned to the surface. 

A typical free energy surface computed from Eq.~\ref{eq:F_gammaL}
is illustrated in Figure $\ref{fig:Two-Dimensional-Model}$c, with
the three equilibrium states satisfying Eq. (\ref{eq:force_balance})
illustrated as red, yellow and black points. The minimum (red point)
corresponds to the stable wetting geometry of the nanowire. Note,
that it is symmetrical, i.e., $\alpha=0$, with contact angles that
are determined by $V_{l}$, $\left(\frac{\gamma_{lv}}{\gamma_{vs}}\right)$
and $\left(\frac{\gamma_{sl}}{\gamma_{vs}}\right)$. These equilibrium
angles are generally different from contact angles $\phi_{vs}^{Young}$
and $\phi_{lv}^{Young}$ for an unconstrained triple line given by
Young's condition: $\gamma_{sl}/\sin\left(\phi_{sl}^{Young}\right)=\gamma_{vs}/\sin(\phi_{vs}^{Young})=\gamma_{lv}/\sin(\phi_{lv}^{Young})$.
In the nanowire geometry, for a given set of surface tensions, Young's
condition appears at a critical liquid volume $V_{l}^{Young}\left(\left(\frac{\gamma_{lv}}{\gamma_{vs}}\right),\left(\frac{\gamma_{sl}}{\gamma_{vs}}\right)\right)$.

The black point in Figure \ref{fig:Two-Dimensional-Model}c corresponds
to the symmetrical ($\alpha=0$) unstable wetting configurations,
while the yellow sphere represents a saddle point configuration with
$\alpha\neq0$. This {}``tilted'' state gives the lowest free energy
barrier for transition from the symmetric to asymmetric configurations,
while the symmetrical transition path goes through the highest energy
barrier and therefore it is the least likely path. The shape of the
free energy surface depends on the droplet volume. As $V_{l}$ increases
the saddle point moves towards the minimum. At a critical volume $V_{l}^{Young}$
the minimum and saddle points merge such that the Hessian of $F$
becomes zero (c.f., Figure~\ref{fig:Two-Dimensional-Model}d) and
the contact angles of the saddle configuration satisfy Young condition.
At $V_{l}>V_{l}^{Young}$ no stable symmetric configuration exists
and the system will undergo unstable evolution towards a tilted configuration.
The value of the liquid volume giving rise to Young's angles can be
calculated knowing $\left(\frac{\gamma_{lv}}{\gamma_{vs}}\right)$
and $\left(\frac{\gamma_{sl}}{\gamma_{vs}}\right)$. Thus, the function
$V_{l}^{Young}=V_{l}^{Young}\left(\left(\frac{\gamma_{lv}}{\gamma_{vs}}\right),\left(\frac{\gamma_{sl}}{\gamma_{vs}}\right)\right)$
determines the stability limits: symmetrical wetting configurations
of nanowires exist for droplet volumes $V_{l}<V_{l}^{Young}$, while
they are unstable for $V_{l}>V_{l}^{Young}$. Changes in nanowire
growth (i.e., environmental) conditions such as temperature, pressure
or composition that modify surface free energies $\left(\frac{\gamma_{lv}}{\gamma_{vs}}\right)$
and $\left(\frac{\gamma_{sl}}{\gamma_{vs}}\right)$, in turn change
$V_{l}^{Young}$. Thus, stable symmetrical nanowires, can become unstable
and tilt upon such sudden changes in environmental conditions.

\begin{figure}
\begin{centering}
\includegraphics[width=0.7\textwidth]{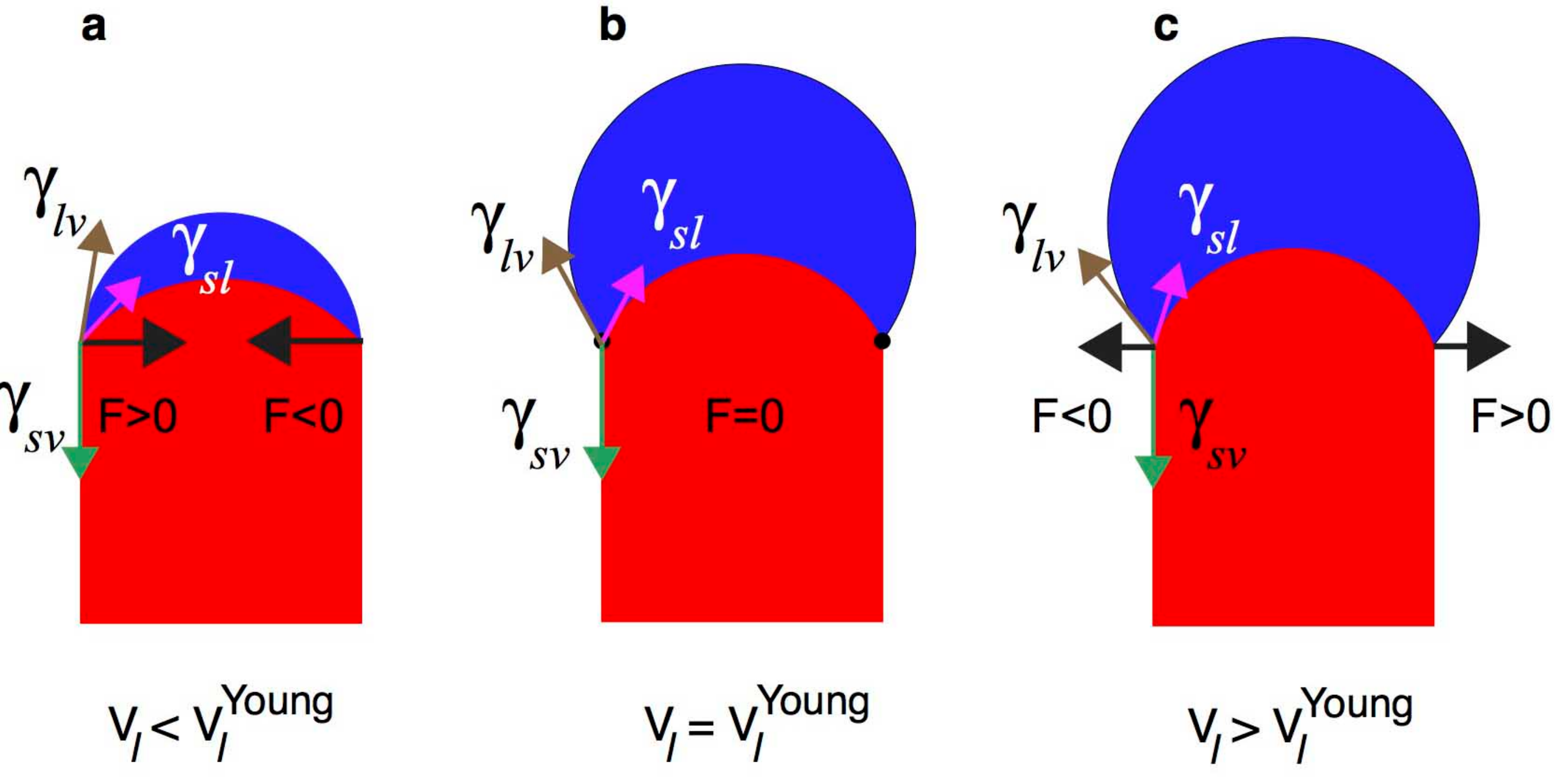}
\par\end{centering}

\caption{\textbf{Capillary forces on triple junctions.} The net capillary forces
on triple junctions for different volumes of the droplet are shown
by black arrows. These are result of the capillary forces from each
of the three interfaces. \textbf{a,} Capillary force is directed normal
to the nanowire axis and points inward towards the nanowire for $V_{l}<V_{l}^{Young}$.
\textbf{b,} Zero force at Young's conditions. \textbf{c,} Capillary
force is directed outward for $V_{l}>V_{l}^{Young}$. In case \textbf{a}
capillary forces push triple junctions towards each other. Since tilting
would increase the separation, it is not favorable. On the other hand,
in case \textbf{c} a small perturbation from a symmetrical configuration
will result in increasing tilting, since capillary forces try to pull
the two triple junctions apart. \label{fig:Force_argument} }
\end{figure}

To interpret the role of the Young condition in determining the stability
of symmetric nanowire wetting geometries we propose the following
capillary force argument. For liquid droplet volumes $V_{l}<V_{l}^{Young}$
the net capillary forces \emph{on} the vapor-solid-liquid triple junction
(TJ) are normal to the sidewall of the nanowire, pointing inwards
toward the solid putting the triple line in tension--the influence
would be towards shortening the distance between the two triple points(c.f.,
Figure \ref{fig:Force_argument}a). As a result any tilting perturbation
is unfavorable as it leads to an increase in the distance between
the TJs. At $V_{l}=V_{l}^{Young}$ the capillary forces are identically
zero (c.f., Figure~\ref{fig:Force_argument}b). Finally, at $V_{l}>V_{l}^{Young}$
the capillary forces point surface away from the solid and the triple
line is in compression: the triple junctions are pushed apart via
tilting (c.f.,Figure~\ref{fig:Force_argument}c). This interpretation
suggests an intrinsic link between the Young's force-balance condition
and the stability of symmetric nanowire wetting geometries, which
is expected to be more generally applicable, beyond the simple model
considered above. Indeed, the correspondence between the predictions
of the model and the MD observations for a system featuring pronounced
interfacial anisotropies, suggest that the theoretical predictions
are applicable to three dimensional systems with anisotropies characteristic
of real VLS systems.

In summary, we have presented a theoretical analysis of equilibrium
wetting geometries on single-crystal faceted solid nanowires, motivated
by observations derived from three-dimensional MD simulations. The
results suggest a tilting instability that is intrinsic to such geometries
under conditions where the solid-liquid interface is deformable, while
the solid/vapor nanowire surfaces remain flat. The transition from
stable to unstable symmetric wetting geometries is shown to be governed
by Young's condition and coincides with a change in the sign of the
resultant force at the triple junction. This simple criterion links
the stability of symmetric wetting geometries to the equilibrium contact
angles that can be varied through changes in growth conditions such
as temperature and partial pressure. It is suggested that the capillary
instability described here is directly related to kinking processes
observed experimentally in VLS nanowire growth, and that the Young's
criteria discussed above provides a relatively simple guideline for
choosing the experimental conditions required to realize straight
versus kinked nanowire shapes.

\section*{Acknowledgments }

This research was supported in part by the US National Science Foundation
under Grant No. DMR-1105409. Use was made of computational resources
provided under the Extreme Science and Engineering Discovery Environment
(XSEDE), which is supported by the National Science Foundation under
grant number OCI-1053575. T.F. was supported for part of this work
by a post-doctoral fellowship from the Miller Institute for Basic
Research in Science at University of California, Berkeley. W.C.C.'s
effort was supported by a visiting Miller Faculty Fellowship, also
from the Miller Institute for Basic Research in Science, during a
sabbatical at the University of California, Berkeley.

\section*{Methods\label{sec:Results and Discussion}}

Molecular dynamics simulations were performed using the LAMMPS software
package.\cite{Plimpton95} To model nanowires wetted by liquid we
used the angularly dependent Stillinger-Weber potential for Si.\cite{Stillinger85}
First we prepared a simulation block with liquid and diamond-cubic
solid phases separated by (111) solid-liquid interface. The interface
plane was normal to the $z$ direction of the simulation block, while
the $x$ and $y$ directions were parallel to crystallographic directions
$[1\overline{1}0]$ and $[11\overline{2}]$ of the solid phase, respectively.
The dimensions of the simulation block were $30\times50\times30$
nm$^{3}$. The phases were equilibrated in a microcanonical (NVE)
ensemble for 5 ns leading to a coexistence temperature of around $1682$
K.

This simulation block was then used to produce input configurations
for nanowire-liquid simulations. Specifically, we carved out columns
with hexagonal crossection and $\{11\overline{2}\}$ side facets containing
solid and liquid phases. The sizes of the solid surface facets ranged
from $5$ to $12$ nm. To investigate the effects of the droplet size
on equilibrium configuration, we varied the amount of liquid phase.
The columns were then equilibrated during a 5 ns long simulation in
a microcanonical . We performed up to 30 ns long simulations of equilibrated
NW configurations. In equilibrium the amounts of solid and liquid
phases do not change and the temperature fluctuates around equilibrium
value.

\begin{figure}
\begin{centering}
\includegraphics[width=0.7\textwidth]{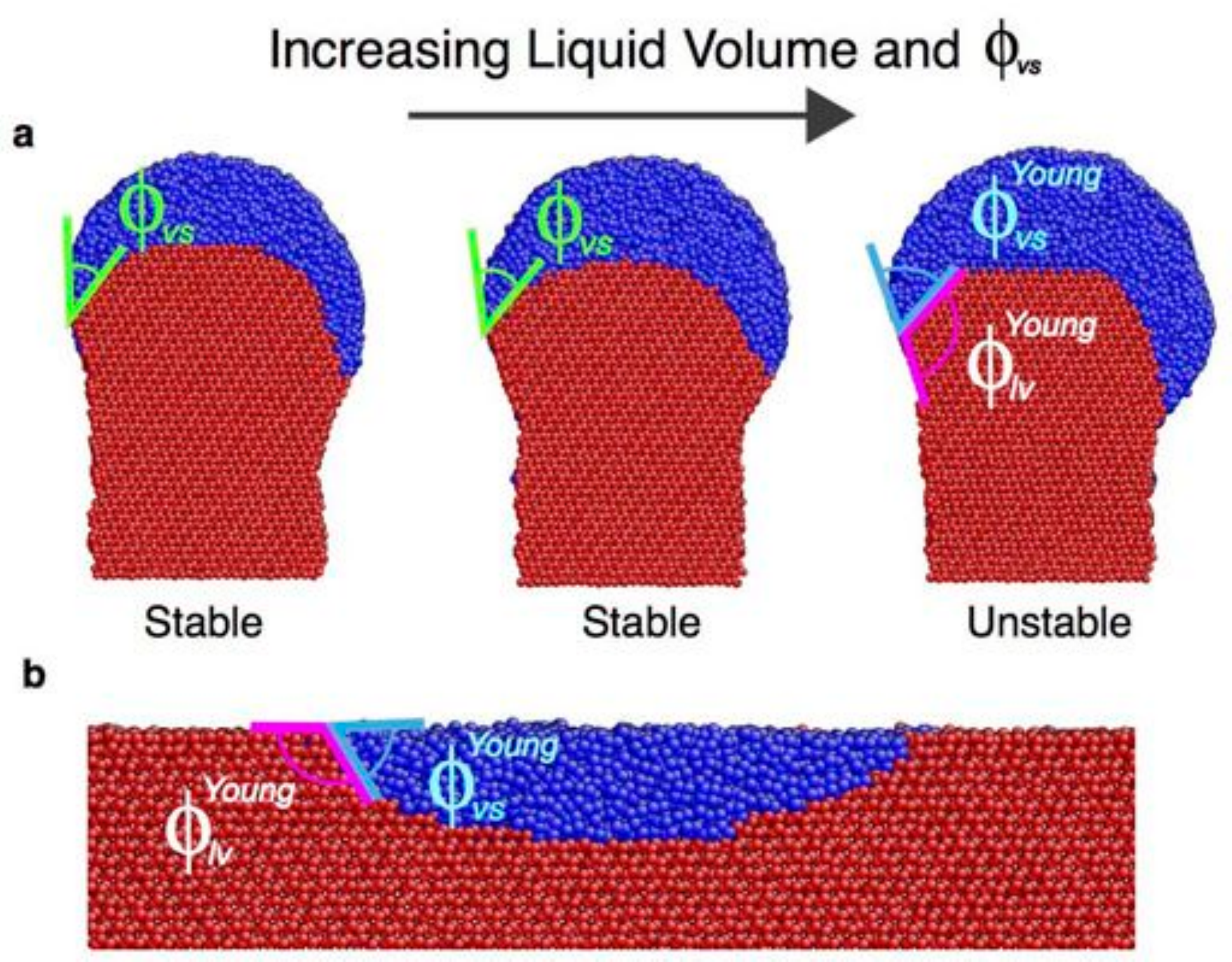}
\par\end{centering}

\caption{\textbf{Nanowires with increasing volume of the liquid droplet.} \textbf{a,}
Equilibrium angles depend on the droplet volume, with $\phi_{vs}$
increasing with $V_{l}$. \textbf{b,} Equilibrium wetting angles of
a liquid droplet on $[111]$ surface. Spontaneous tilting occurs when
nanowire contact angles approach angles indicated in \textbf{b}.\label{fig:Kinking_MD}}
\end{figure}

Equilibrium simulations for an isolated droplet on a \{111\} solid
surface, shown in Figure \ref{fig:Kinking_MD}b were performed in
a microcanonical ensemble for 10 ns.\cite{Frolov11a} The solid slab
used in these simulations had dimensions $2.3\times6.7\times32.9$
nm$^{3}$ parallel to $[1\overline{1}0]$, $[111]$ and $[11\overline{2}]$
crystallographic directions, respectively.


\bigskip{}

\pagebreak{} 

\newpage{}

\clearpage{}

\clearpage{} 
\end{document}